\documentstyle[aps,prb,epsfig,twocolumn]{revtex}
\begin{document}
\newcommand{\vektor}[1]{\mbox{\boldmath $#1$}}
\newcommand{\taubold}{\mbox{\boldmath$\tau$}}
\twocolumn[\hsize\textwidth\columnwidth\hsize\csname
@twocolumnfalse\endcsname
\title{Localized spin ordering in Kondo lattice models}
\author{I. P. McCulloch${}^{\dagger}$,
A. Juozapavicius${}^{\ddagger}$, A. Rosengren${}^{\ddagger}$
and M. Gulacsi${}^{\dagger}$}
\address{${}^{\dagger}$Department of Theoretical Physics,
Institute of Advanced Studies, \\
The Australian National University,
Canberra, ACT 0200, Australia}
\address{${}^{\ddagger}$Department of Theoretical Physics,
Royal Institute of Technology, \\
SE-100 44 Stockholm, Sweden}
\date{\today} 
\maketitle
\begin{abstract}
Using a non-Abelian density matrix renormalization
group method we determine the phase diagram of the Kondo lattice model
in one dimension, by directly measuring the magnetization of the
ground-state. This allowed us to discover a second ferromagnetic phase
missed in previous approaches. The phase transitions are found to be
continuous. The spin-spin correlation function is studied in detail,
and we determine in which regions the large and small Fermi surfaces
dominate. The importance of double-exchange ordering and its
competition with Kondo singlet formation is emphasized in
understanding the complexity of the model.
\end{abstract}
\pacs{PACS No. 75.50 Mb, 71.10 Hf, 05.10.Cc} 
]

The Kondo lattice model (KLM) describes the interaction between
a conduction electron (CE) band and a half-filled narrow impurity,
e.g. $f$ electron, band and is thought to capture the essential
physics of the rare earth compounds. Although intensively studied
for two decades, the KLM is still far from being completely
understood. Recently, after the discovery of Kondo insulators
and the non-Fermi liquid behavior, interest in this field
has been greatly renewed, especially due to the non-Fermi
liquid behavior discovered in most of the heavy fermion compounds,
which resembles a Griffiths phase.\cite{neto}

In order to understand the role of the impurity spin in
determining the properties of KLM we must develop a better
understanding of the magnetic correlations.
The Griffiths phase in the one dimensional
KLM occurs naturally;\cite{graeme} it is, therefore, the
prototypical model for heavy fermion compounds.
Hence, this is an ideal system to study since we
have the bosonized solution \cite{graeme,boso} and we
know the behavior of the CEs  in both the paramagnetic (PM)
and ferromagnetic (FM) phases. However, less attention has
been given to understand the correlations between the
impurity spins. This is the focus of our study.

The Hamiltonian for the KLM is
\begin{equation}
H \: = \: -t \sum_{j=1, \sigma}^{L} ( c^{\dagger}_{j, \sigma}
c^{}_{j + 1, \sigma} + {\rm h.c.} ) \: + \:
J \sum_{j=1}^{L} {\bf S}^{c}_{j} {\bf \cdot} {\bf S}^{}_{j} \; ,
\label{klm}
\end{equation}
where $t > 0$ is the CE hopping parameter,
${\bf S}_{j}$ are spin $1/2$ operators for the
localized spins, e.g. $f$, and
${\bf S}^{c}_{j}= \frac{1}{2} \sum_{\sigma^{},\sigma^{\prime}}
c^{\dagger}_{j, \sigma^{}} {\vektor{\sigma}}_{\sigma^{},\sigma^{\prime}}
c^{}_{j, \sigma^{\prime}} $
with ${\vektor{\sigma}}$ the Pauli spin matrices and
$c^{}_{j, \sigma}$, $c^{\dagger}_{j, \sigma}$ the electron
annihilation and creation site operators. The Kondo coupling
$J$ is measured in units of the hopping $t$ and partial conduction
band filling, $n = N/L < 1$, is assumed throughout.

The method that we use is density matrix renormalization group (DMRG)
which, however, is extended to explicitly preserve $SU(2)$ spin and
pseudospin symmetry. Hence we can measure the magnetization directly
and determine rigorously the PM - FM phase boundary. The obtained
result is in excellent agreement with a recent bosonized solution
\cite{graeme} and contradicts common view that this phase boundary
goes to infinite Kondo coupling $J$ as the CE density approaches
half-filling.\cite{RevModPhys,JPhys} We also determine the regions of
the phase diagram where large and small Fermi surfaces are dominant,
which has been a central issue for much of the research in this area
for some years.

In addition, we have discovered a second FM region
not seen before. For most dopings, this region of FM
separates the regions of large and small Fermi surface. This
most likely resolves the question as to the applicability
of the Luttinger theorem to the KLM, shown by Yamanaka {\it et al.},
\cite{Luttinger} since the Fermi points are not expected to
remain constant across a phase transition. 

To accelerate the computation, we make use of several operators that
commute with the Hamiltonian, $S^+, S^-, S^z, I^+, I^-, I^z$,
respectively the generators of the spin $SU(2)$ and pseudospin $SU(2)$
algebras.\cite{PseudoSpin} Combined, the generators form the algebra
$SO(4)$. All of the states in our DMRG calculation transform as
irreducible representations of this algebra. Since $SO(4)$ is
non-Abelian these representations have, in general, degree $> 1$,
which implies that a single basis state in the $SO(4)$ representation
is equivalent to multiple states of the purely Abelian representation
of most previous DMRG calculations. This is the origin of the dramatic
performance improvements of the non-Abelian DMRG.  The states are
labeled by the eigenvalues of the Casimir operators of $SO(4)$, which
are $S^2 = s(s+1)$ and $I^2 = i(i+1)$. Hence we can label all
irreducible representations by $[s,i]$, which has degree
$(2s+1)(2i+1)$. In this construction, a chemical potential would appear
as a term in the Hamiltonian proportional to $I^z$, acting in an
identical way a magnetic field coupled to $S^z$.
Although the basis states in the calculation are eigenstates of
$S^2$ and $I^2$, rather than $S^z$ and $I^z$, all these operators
mutually commute so it is possible to simply replace $S^z$ and
$I^z$ by the chosen eigenvalues in this case.
A single site of the Kondo lattice contains just three
such states. The simplest is the Kondo singlet state, transforming as
the $[0,0]$ representation of degree 1. The Kondo triplet state
transforms as the $[1,0]$ representation of degree 3, and encapsulates
the three projections $\vert \Uparrow \uparrow \rangle$, $\sqrt{1/2} (
\vert \Uparrow \downarrow \rangle + \vert \Downarrow \uparrow
\rangle)$, $\vert \Downarrow \downarrow \rangle$ in a single
state. Here, $\Uparrow$ denotes localized $f$, and $\downarrow$ the
conduction electron spins, respectively. Finally, the holon state
(actually, the tensor product of a holon and a $f$ spin) transforms as
the $[1/2,1/2]$ representation of degree 4 and has the projections
$\vert \Uparrow 0 \rangle$, $\vert \Downarrow 0 \rangle$, $\vert
\Uparrow \uparrow \downarrow \rangle$, $\vert \Downarrow \uparrow
\downarrow \rangle$.  The single-site operators are $3 \times 3$
matrices over this basis.  The matrix elements can be determined by
the Wigner-Eckart theorem, which specifies the relationship between
the 3 dimensional reduced basis and the full 8 dimensional basis. For
a comprehensive description of the new algorithm, see
Ref.\ \onlinecite{totalspinDMRG}. At half filling (where the ground state is
a pseudospin singlet) 400 block states are equivalent to around 2500
states of a calculation using $N$ and $S^z$ quantum numbers, although
the relative advantage of $SO(4)$ decreases as the system is doped
away from half filling. We used the new DMRG algorithm to obtain the
ground state energy, magnetization and different correlation
functions, i.e., the momentum distribution, density-density,
conduction electron spin-spin and the $f$ spin structure factor,
$S(k)$. The obtained results can be summarized with the phase diagram
presented in Fig.\ 1, which will be analyzed in detail hereafter. The
main properties of the phase diagram have been confirmed on chains of
120 or more sites. Results for the magnetization were calculated on
smaller chains, 40 - 60 sites, where the energies can be calculated
more accurately.  We found no finite size effects that would
affect the properties of Fig.\ 1. In all cases, we extrapolate
to zero truncation error based on well-converged sweeps of
between 200 and 500 $SO(4)$ states kept.

As it can be seen from Fig.\ 1, the main feature dominating the KLM is
$f$ spin FM ordering. The FM ordering is due to the
double-exchange (DE) interaction which appears as a consequence
of an excess of localized spins over CEs: \cite{zener}
each CE has to screen more than one localized spin, and
since hopping is energetically most favorable for CEs
which preserve their spin, this tends to align the localized
spins. This element was missing in the early approaches,
which concentrated on the competition between
Kondo singlet formation at large $J$ and the RKKY interaction
in the weak coupling limit.\cite{Jullien} This picture
is borrowed from the single impurity Kondo model and is
inadequate for the lattice case.\cite{RevModPhys,JPhys} 

Starting the analysis of the phase diagram for large $J$,
we see that all CEs form singlets with the localized
$f$ spins \cite{largeJ} which become inert. The uncoupled $f$ spins
order FM in a mechanism similar to the $J < 0$ case.\cite{zener}
Here, there is no competition between Kondo singlet
formation and DE. The fully polarized state [with $S = (L-N)/2$]
appears for any value of $n < 1$ \cite{graeme,largeJ}
contrary to the suggestion of Refs.\ \onlinecite{RevModPhys,JPhys}
that close to half filling the PM phase extends to $J \rightarrow \infty$.
As $J$ is lowered, KLM can be rigorously mapped into a random transverse
field Ising model,\cite{graeme} hence the phase transition (the
solid curve in Fig.\ 1) is identical to the quantum order - disorder
transition. It should be emphasized that this is also true for
the second FM phase, as will be shown later on.

%%%%%%%%%%%%%%%%%%%%%%%%%%%%%%%%%%%%%%%%%%%%%%%%%%%%%%%%%%%%%%%%%%%%

\begin{figure}[h]
\epsfxsize=7cm
\centerline{\epsfbox{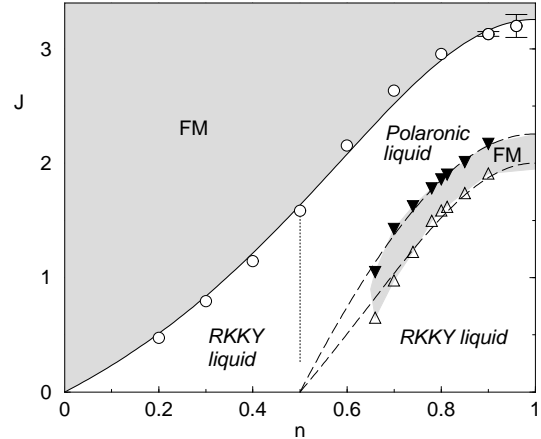}}
\caption{
The obtained phase diagram of KLM. The two shaded areas are
the FM phases. The open circles and triangles correspond to
points at which the FM energy level crosses the $S=0$ level.
The dashed curves are the derived phase transition lines
(the solid curve was already obtained in
Ref.\ \protect\onlinecite{graeme}).}
\label{fig1}
\end{figure}

%%%%%%%%%%%%%%%%%%%%%%%%%%%%%%%%%%%%%%%%%%%%%%%%%%%%%%%%%%%%%%%%%%%%
%%%%%%%%%%%%%%%%%%%%%%%%%%%%%%%%%%%%%%%%%%%%%%%%%%%%%%%%%%%%%%%%%%%%

\begin{figure}[h]
\epsfxsize=7cm
\centerline{\epsfbox{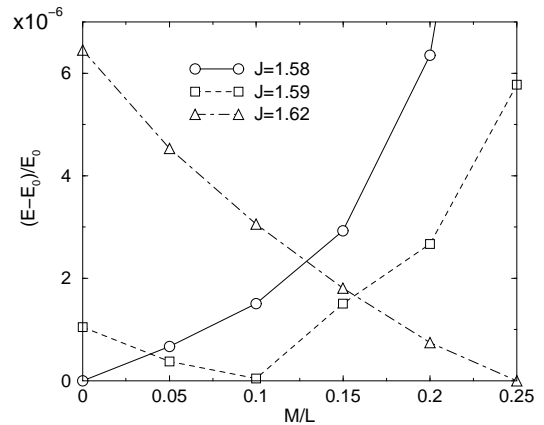}}
\caption{
Normalized magnetization curves (relative to the
ground state energy, $E_0$) across the phase transition
at quarter filling for a 40 site lattice.}
\label{fig2}
\end{figure}

%%%%%%%%%%%%%%%%%%%%%%%%%%%%%%%%%%%%%%%%%%%%%%%%%%%%%%%%%%%%%%%%%%%%

The phase transition obtained via DMRG fits exceptionally well this
picture, confirming the bosonization result of Ref.\ \onlinecite{graeme}.
The open circles correspond to points at which the energy of the FM
state crosses the energy of the singlet state. Since the phase
transition is second order, this is only an upper bound on the
true transition line. However the partially polarized region is
very small, of the order of $J/t \sim 0.01$, which is why
this phase transition has not previously been observed to be
continuous. A typical example of the energy versus the
magnetization ($M$) is presented in Fig.\ 2. This shows that in
the transition regime, $\partial^2 E / \partial M^2$ is positive.
We have accounted for all known random errors, these are errors
arising from the tolerance of the matrix diagonalization,
variations in the energy across the DMRG sweep, and error
arising from the extrapolation to zero truncation error.
These errors are of the order of the symbol size in this figure.

Below the solid curve, Fig.\ 1, the Kondo singlets are not
inert anymore and they greatly contribute to the properties of
KLM. Excluding the Kondo triplet states, the CE wave function in the
continuum limit satisfies a nonlinear Schr\"{o}dinger equation
\cite{polaron} which has finitely delocalized solitonic solutions.
\cite{holstein} This corresponds to a dressing of the CE by a cloud
of antiparallel local spins, i.e., spin polarons are formed. The
polaronic length scale competes with the length scale set by the free
CE mean free path and introduces competing time scales: slow motion of
the polarons with low energy dynamics and fast motion of the free CEs
with high energies. This scenario resembles a two-fluid picture with
intrinsic inhomogeneities which involves spin fluctuations and
short-range spin correlations, which we call a {\sl polaronic liquid}.

%%%%%%%%%%%%%%%%%%%%%%%%%%%%%%%%%%%%%%%%%%%%%%%%%%%%%%%%%%%%%%%%%%%%

\begin{figure}[h]
\epsfxsize=7cm
\centerline{\epsfbox{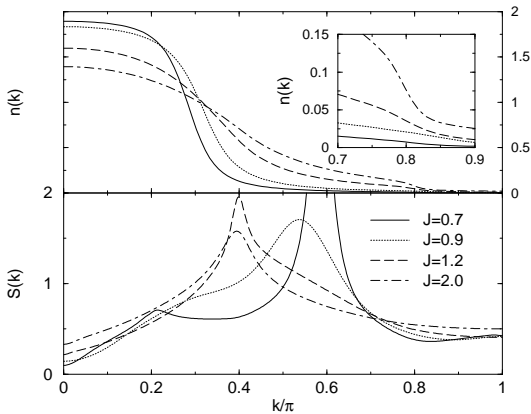}}
\caption{
Typical $J$ dependence of the spin structure factor, $S(k)$,
and the momentum distribution, $n(k)$ ($n = 0.6$).}
\label{fig3}
\end{figure}

%%%%%%%%%%%%%%%%%%%%%%%%%%%%%%%%%%%%%%%%%%%%%%%%%%%%%%%%%%%%%%%%%%%%

Finite temperature DMRG \cite{shibata} confirmed the presence
of short-range $f$ spin correlations in the van-Hove singularities.
Consequently the structure factor peaks at $2 k_F - \pi$,
where $k_F$ is the Fermi point determined by the filling of the CE
band. This means that the localized $f$ spins, even though they
are completely immobile, contribute to the volume of the Fermi sea.
This conventionally is called a {\sl large} Fermi surface, the
effect of which is also seen in the momentum distribution
function, see Fig.\ 3. As the polarons are formed the peak
of $S(k)$ shifts from the small $J/t$ value of $2 k_F$:
the slow motion of the spin polarons will dominate the
low energy dynamics of the quasiparticles. This proves
that the appearance of the large Fermi surface is a
dynamical effect since it involves local inhomogeneities,
impurity spin fluctuation, and short-range correlations
of the $f$ spins. This phase is related \cite{graeme} to a
Griffiths phase, suggesting that the small - large Fermi
surface crossover is a Griffiths singularity.

The large Fermi surface is conventionally explained by
reference to the periodic Anderson model (PAM) ancestry.
\cite{JPhys,Luttinger,largeFS} Our results imply that
even for PAM, this simple picture is inadequate. In
particular, we see no reason why a small - large
Fermi surface crossover, marked by a FM phase,
should not also appear in PAM. However, the
behavior of the Fermi surface crossover close to quarter
filling is numerically difficult to determine (dotted
line in Fig.\ 1), hence we are not yet able to rule
out the possibility that the large and small Fermi
surface regions are adiabatically connected. Even prior
to the current calculation, the nature of the Fermi surface 
in the weak-coupling regime was not clear, with the suggestion 
from Ref.\ \cite{JPhys} that the Fermi surface vanishes at 
a point in proximity to where we find the ferromagnetic phase.
For $n < 0.5$
the width of the polarons is over several lattice spacings
(diverging for $n \rightarrow 0$, \cite{polaron}) hence the
energy needed to excite these polarons is too large for this
effect to happen. The polarons will not contribute to the low
energy dynamics and the system behaves as an RKKY liquid, as we
explain below. 

An interesting phenomenon appears as we further lower $J$.
The residual weight attached to the Kondo singlets vanishes,
hence all CEs which participated in the formation of these
singlets, become delocalized. The distance between these CEs
is much larger than the lattice spacing, and below
$J \le 2 {\sqrt{n}} \sin ( \pi n )$ their continuum
limit takes the regular quantum sine-Gordon form.\cite{boso}
In the bosonization language of Ref.\ \onlinecite{graeme}, this
means that the spin Bose fields, $\Phi_{\sigma}$ cannot be
approximated by their noninteracting expectation values,
rather by their expectation value corresponding to a
sine-Gordon (sG) model, $\Phi_{\sigma} \approx \langle
\Phi_{\sigma} \rangle_{\rm{sG}}$. However, the charge degrees
of freedom not being affected by the sine-Gordon spin gap,
their corresponding Bose fields, $\Phi_{\rho}$ may be still
approximated by their noninteracting values. Extending the
bosonized results of Ref.\ \onlinecite{graeme} to a finite
$\langle \Phi_{\sigma} \rangle_{\rm{sG}}$, we obtain
the critical Hamiltonian governing the PM - FM
phase transition at intermediate $J$ values in the form:
$H_{\rm crit.} = - J^2 {\cal{A}} / (2 \pi^2 v_F) \sum_{j} 
{\bf S}^{z}_{j} {\bf S}^{z}_{j + 1} + 2 J {\cal{B}} \sum_{j} 
\{ 1 - ( \langle \Phi_{\sigma} \rangle^{2}_{\rm{sG}} / 2 ) 
[1 + J / (2 \pi v_F) ]^2 + \cos ( 2 k_F j ) \} {\bf S}^{x}_{j}$,
where ${\cal{A}}$ and ${\cal{B}}$ are functions which depend only
on the cutoffs introduced by the bosonization
scheme.\cite{graeme,boso} Following closely previous
bosonization approaches,\cite{graeme,boso} we can prove that
the critical behavior of the FM transition for the intermediate
this $J$ case is of a random transverse-field Ising
model type, where the transverse field
$h_j = 2 J {\cal{B}} \{ 1 - ( \langle \Phi_{\sigma} \rangle^{2}_{\rm{sG}} / 2 ) 
[1 + J / (2 \pi v_F) ]^2 + \cos ( 2 k_F j ) \}$
is driven by a displaced cosine distribution
of the form: 
$\rho (h) = [1 / (2 \pi J {\cal{B}} ) ] \{ 1 - [ h / (2 J {\cal{B}} ) + 
( \langle \Phi_{\sigma} \rangle^{2}_{\rm{sG}} / 2 ) 
[1 + J / (2 \pi v_F) ]^2 - 1 ]^2 \}^{-1/2}$.
Accordingly, the FM transitions emerging at
intermediate values of $J$ are of a quantum order -
disorder type. These transitions are driven by spin polarons,
contrary to the FM phase emerging at high $J$ values, which
is given by the uncoupled $f$ spins (in a mechanism similar
to the $J < 0$ case).  The new critical line is 
$J_c = \alpha ({\cal {A}}, {\cal {B}}) \sin (\pi n /2) / 
[1 - \beta ({\cal {A}}, {\cal {B}})] - \gamma ({\cal {A}}, {\cal {B}}, 
\langle \Phi_{\sigma} \rangle^{2}_{\rm{sG}})$.
The bosonization (conformal field-theory) arguments
does not determine the magnitude of $\alpha$, $\beta$
and $\gamma$, accordingly these constants are used as
fitting parameters to the numerically obtained points.
The best fits are the dashed curves in Fig.\ 1.

This is the second FM phase in Fig.\ 1, which has proven difficult to
detect with conventional (Abelian) DMRG.\cite{exactFM,numerics} Previous
DMRG calculations did show a weak FM signal at $n = 0.8$ and $J = 1.6$
and $1.8$,\cite{largeFS} but the results were discarded in later papers
by the same authors.\cite{JPhys,RevModPhys} Likewise an
exact diagonalization of a very small system gave FM for $n=0.75$ and
$J=1.5$.\cite{exactFM} Using the non-Abelian DMRG algorithm we could
also check the energy of each total spin state, shown in Fig.\ 4,
which clearly shows a second ferromagnetic region although we have
not yet confirmed numerically the order of the phase boundaries.
For the FM Kondo lattice model, $J < 0$, a phase separated regime
was observed in numerical approaches.\cite{dagotto} However for
$J > 0$ we found no change in sign of the inverse charge
compressibility. Thus, this phase is a true FM rather than a
phase separated region. 

%%%%%%%%%%%%%%%%%%%%%%%%%%%%%%%%%%%%%%%%%%%%%%%%%%%%%%%%%%%%%%%%%%%%

\begin{figure}[h]
\epsfxsize=7cm
\centerline{\epsfbox{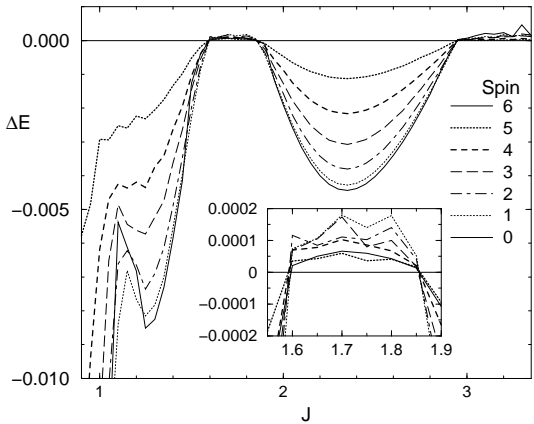}}
\caption{
The gap, $\Delta E$, from the fully polarized ferromagnetic
state to every other spin state {\it vs.} $J$,  for $n=0.8$,
and 60 site lattice, measured along intervals of $J \pm 0.05$.
For most data points the error bars are of order $\sigma_{\Delta E}
\sim 10^{-5}$ or less, except for the $S=0$ curve for very low
and very high $J$, where the errors are of order
$\sigma_{\Delta E} \sim 5 \times 10^{-4}$. The inset 
shows the second ferromagnetic region.}
\label{fig4}
\end{figure}

%%%%%%%%%%%%%%%%%%%%%%%%%%%%%%%%%%%%%%%%%%%%%%%%%%%%%%%%%%%%%%%%%%%%

Below the second FM region the KLM reduces to a system of free
localized spins in fields determined by CE scattering: dominant
$2k_{F}$ modulations are manifest, see Fig.\ 3, superimposed on
an incoherent background. This reflects the momentum transferred
from the CE band to the spin chain in backscattering interactions,
together with incoherent forward scattering. This case is referred
to as an RKKY liquid as the scattering processes give an RKKY-like
correlation for the $f$ spins, even though the RKKY interaction
strictly diverges in one dimension.

In conclusion, using a non-Abelian DMRG method a most comprehensive
analysis of the short- and long-range ordering of the localized moments
in KLM is presented.  We show that DE ordering and its competition with
Kondo singlet formation is the dominant feature of the phase diagram.
The non-Abelian DMRG method allowed us to discover that FM does not
only appear at large $J$ but also at intermediate values. This second
FM phase was missed in previous approaches. We also show that at large $J$
FM is due to ordering of uncoupled $f$ spins, while for intermediate $J$,
i.e., the second FM region, FM is due to ordering of the spin polarons.
The inhomogeneous polaronic state between these two FM phases
is analogous to a two-fluid system and it exhibits a large Fermi
surface.

Work in Australia was supported by the Australian Research Council. In
Sweden by the Swedish Natural Science Research Council.

\end{document}